\documentclass[pra,aps,floatfix,twocolumn,showpacs,reprint,superscriptaddress,linenumbers]{revtex4}
\usepackage{lipsum}
\usepackage{subfigure}
\usepackage{hyperref} 
% ...
\usepackage[T1]{fontenc}             
\usepackage[applemac]{inputenc} 
\usepackage{mathptmx}                
\usepackage[scaled=.90]{helvet}
\usepackage{graphicx}                
\usepackage{color}                
\usepackage{setspace}                
\usepackage[sort&compress]{natbib}   
\usepackage[loose,nice]{units}       
\usepackage{tabularx}                
\usepackage{booktabs}                
\usepackage{amsmath}
\usepackage{ulem}
\usepackage{amssymb}
\usepackage{url}
\newcolumntype{C}{>{\centering\arraybackslash}X}
\newcolumntype{L}{>{\raggedleft\arraybackslash}X}
\newcolumntype{R}{>{\raggedright\arraybackslash}X}

\usepackage{graphicx}
\usepackage{textgreek}
\usepackage{sidecap}
\usepackage{floatrow}
\usepackage{sidecap}  
\usepackage{dcolumn}
\usepackage{bm}
\usepackage{hyperref}%
\hypersetup{
    colorlinks=true,
    linkcolor=blue,
    citecolor=blue,
    filecolor=magenta,      
    urlcolor=blue,
}

\begin{document}


\date{\today}

\title[]{Generation of Super Intense Isolated Attosecond Pulses from Trapped Electrons in Metal Surfaces}

\author{Younes Adnani}
\affiliation{University Ibn Tofa\"{i}l, High School of Technology, Laboratory of Electronic Systems, Information Processing, Mechanics and Energy (SETIME),Kenitra, Morocco}
\author{Abdelmalek Taoutioui}
\affiliation{Institute for Nuclear Research, ATOMKI, Debrecen, Hungary}
\author{Abdelkader Makhoute}
\affiliation{Faculty of Sciences, Free University of Brussels (ULB), Campus de la Plaine, 1050 Brussels- Belgium}
\affiliation{Dynamics of Complex systems, Faculty of Sciences, Moulay Ismail University, Meknes- Morocco}
\author{K\'{a}roly T\H{o}k\'{e}si}
\affiliation{Institute for Nuclear Research, ATOMKI, Debrecen, Hungary}
\author{Hicham Agueny}
\email{hicham.agueny@uib.no}
\affiliation{Department of Physics and Technology, Allegt. 55,
University of Bergen, N-5007 Bergen, Norway 
}%
\begin{abstract}
Generation of ultrabroadband isolated attosecond pulses (IAPs) is essential for time-resolved applications in chemical and material sciences, as they have the potential to access the spectral water window region of chemical elements, which yet has to be established. Here we propose a numerical scheme for highly efficient high-order harmonic generation (HHG) and hence the generation of ultrabroadband IAPs in the XUV and soft x-ray regions. The scheme combines the use of chirped pulses with trapped electrons in copper transition-metal surfaces and takes advantage of the characteristic features of an infrared (IR) single-cycle pulse to achieve high conversion efficiencies and large spectral bandwidths. In particular, we show that ultrabroad IAPs with a duration of 370 as and with a bandwidth covering the photon energy range of 50-250 and 350-450 eV can be produced. We further show that introducing an additional IR single cycle pulse permits to enhance the harmonic yield in the soft x-ray photon energy region by almost 7 order of magnitude. Our findings thus elucidate the relevance of trapped electrons in metal surfaces for developing stable and highly efficient attosecond light sources in compact solid-state devices.
\end{abstract}


\maketitle

\section{INTRODUCTION}\label{intro}
Recent advances of ultrafast laser technology have made it possible to produce  coherent extreme-ultraviolet (XUV) radiation and attosecond pulses~\cite{Krausz2001,Corkum2007,Krausz2009,Krausz2014,Li2018}, which have allowed in a pump-probe experiment to access the ultrafast microscopic processes in atoms, molecules and solids. This fast progress has led to the emerging field of attosecond nanoscopy~\cite{Ghimire2011,Vampa2015a,Ciappina2017,Vampa2017,Lakhotia2020}, which promises to open up new routes for attosecond technology, in particular, petahertz electronics~\cite{Garg2016,Schoetz2019,Yang2020} and attosecond photonics~\cite{Ghimire2019}.

In general XUV lasers and attosecond pulses can be produced by exploiting a coherent and non-linear optical process known as high-order harmonic generation (HHG), first reported three decades ago from atomic gas-phase~\cite{Ferray1988} and now extended to solid-state systems. The HHG process from gas-phase has thus formed the basis of attosecond science, which has now extended to the condensed matter phase, thus offering the possibility of developing compact solid-state HHG devices. The underlying dynamics of the process in gas-phase is well established and understood on the basis of the three-step model~\cite{Corkum1993,Schafer1993,Lewenstein1994}. The model involves tunnel ionization, free acceleration and recombination in the presence of an oscillating electric field. These three aspects of the process are well-known to be the main cause of the emission of high-frequency radiation, which manifests in the HHG spectra by odd harmonics followed by a plateau characterized by the harmonic cutoff photon energy. The latter is proportional to the driving laser intensity and inversely proportional to the square of the driving laser wavelength.

Although extensive works (e.g. ~\cite{Ghimire2011,Zaks2012,Vampa2015a,Vampa2015b,Hohenleutner2015,You2016}) have addressed solid-state HHG, studies of the process at the surfaces and interfaces are not well explored. Here, available studies are limited to HHG from the surface state in Cu(111)~\cite{Aguirre2016,Agueny2021}. At the surface and interfaces, the electrons do not have enough energy to escape into the vacuum nor to penetrate into the bulk as they possess energies within the band gap of a material~\cite{Echenique1986}. Consequently, these electrons get trapped by their image potential, which is created by the polarisation charge induced at the surface~\cite{Echenique1986}. And because of these characteristic features, metal surfaces unlike bulk crystals, can generate electron states classified as intrinsic surface states and image-potential states, and that depend on their charge density localization relative to the surface atomic layer~\cite{Chulkov1999}. As these states are localized at the outermost atomic surface layers, they are easy to be manipulated, which in turn affect the physical and chemical properties of materials, thus making them of particularly importance for nanotechnology and catalysis.

Despite recent achievements in solid-state HHG and in the generation of IAPs, there are still major needs in producing ultrabroad IAPs with higher flux intensities. The generated IAPs have been reported so far with a limited bandwidths and mostly from bulk crystals. For instance, the predicted IAPs in recent theoretical works were found to cover the energy windows (16-20) eV in MoS$_2$ \cite{Guan2020}, (20-35) eV and (17.5-43) eV in bulk MgO \cite{Shao2020,Nourbakhsh2021}, which in general are insufficient for time-resolved measurements on unprecedented time-scales. 

In this work, we demonstrate a nanoscale platform for extending the generation of IAPs to the soft x-ray region (i.e. from 20 eV to 450 eV) and simultaneously increasing their flux intensities by almost 7 order of magnitude. Our scenario is based on exploiting the coherent aspect of the HHG process and makes use of a chirped pulse combined with a single-cycle pulse applied to trapped electrons in metal surfaces Cu(111) and Cu(100). In particular, we show that the predicted IAPs can be extracted with a duration of about 370 as and with a bandwidth covering the photon energy ranges 50-250 eV and 35-450 eV. We further show that the intensity of these pulses can be enhanced by 7 order of magnitude by introducing an IR single-cycle pulse and that the generated IAPs can be controlled by changing the surface orientations Cu(111) and Cu(100). 

A key feature of our work is exploiting the characteristics of trapped electrons in metal surfaces together with tuning the optical properties of chirped pulses combined with a single-cycle pulse. This is shown here to enhance the conversion efficiency of HHG process, unlike conventional schemes applied to periodic systems in which a fundamental multi-cycle pulse is combined with either a few-cycle pulse or a multi-cycle pulse. On the other hand, introducing plasmonic fields has been shown to be efficient for controlling HHG in atomic gases \cite{Ansari2018}. Our work thus provides new insights into the role of trapped electrons in metal surfaces for generating ultrabroad IAPs, which is the key for advancing attosecond science.

This paper is organized as follows. In Sec. \ref{theory} we present our model, which is based on solving the time-dependent Schr\"{o}dinger equations (TDSE) using a one-dimensional (1D)-model as well as its numerical implementation which is followed by the discussion of the results of HHG and their time-frequency analysis [cf. Sec. \ref{results}]. The essence of the study is that IAPs can be produced from image states of metal surfaces with the use of optimized chipred pulses, and their characterization can be achieved via the crystalographic orientation of the metal surface. Atomic units (a.u.), $e=$ $m_e=$ $4\pi \epsilon_0=$ 1, are applied unless otherwise stated.

\section{THEORETICAL AND COMPUTATIONAL MODELS}\label{theory}

The coherent electron motion induced by means of light pulses in transition-metal surfaces is modelled using a 1D-model with the use of an effective pseudo potential as in our previous work~\cite{Agueny2021}. The 1D-model has been introduced in many theoretical works (e.g.~\cite{Aguirre2016,Kazansky2009}), and it was found to provide new insights into the experimental findings~\cite{Kazansky2009}. This minimal model is justified as in a 3D-model, the metal surfaces behave as a quantized 1D system in the direction perpendicular to the surface, and as a free-electron system in the direction parallel to the surface. In addition, we consider a scenario in which a linearly polarized laser pulse is directed to the [111] and [100] directions of the considered metal surface. The TDSE governing this dynamics can be written as 
\begin{equation}\label{tdse}
\Big[-\frac{\nabla_z^{2}}{2} +V_{ion}(z) + H_I(t) - i\frac{\partial}{\partial t}\Big]|\psi(t) \rangle=0,
\end{equation}
where $V_{ion}(z)$ is a one-electron pseudo potential interaction. Here, we use Chulkov potentials~\cite{Chulkov1999} to model the electronic structures of the metal surfaces Cu

\begin{subequations}\label{Vion}
\begin{align*}
V_1(z)&=A_{10} + A_1\cos(\frac{2\pi}{a_s}z), z<0 \\
V_2(z)&=-A_{20} + A_2\cos(\gamma z),  0<z<z_1 \\
V_3(z)&=A_3\exp[-\alpha(z- z_1)],  z_1<z<z_{im} \\
V_4(z)&=\frac{\exp[-\lambda(z- z_{im})]-1}{4(z-z_{im})}, z_{im}<z 
\end{align*}
\end{subequations}
This analytical potential is characterized by the bulk interlayer spacing $a_s$ and the position of the image plane $z_{im}$. It is parameterized with $A_{10}$, $A_1$, $A_2$ and $\beta$, which are independent parameters and are given in the table~\ref{table1} for both Cu(111) and Cu(100). These parameters are obtained by combining density function calculations and experiments~\cite{Chulkov1999}. The other parameters $A_{20}$, $A_3$, $\alpha$, $z_1$, $\lambda$ and $z_{im}$ are determined from the continuity condition of the potential and its first derivative, and are given according to~\cite{Chulkov1999,So2015}   
\begin{eqnarray}
A_{20}=A_2-A_{10}-A_1; \;\; A_3=-A_{20}-\frac{A_2}{\sqrt{2}}\nonumber\\
z_i=\frac{5\pi}{4\gamma}; \;\; \alpha=\frac{A_2\gamma}{A_3}\sin(z_1\gamma)\nonumber\\
\lambda=2\alpha; \;\; z_{im}=z_1-\frac{1}{\alpha}ln(-\frac{\alpha}{2A_3}).
\end{eqnarray}

\begin{table}
\caption{\label{table1}Adjustable parameters used in the analytical pseudo potential in Eq. (\ref{Vion}) and which is shown in Figs. \ref{fig1}(a) and (c). The values are taken from~\cite{Chulkov1999}. All parameters are given in atomic units.}
\begin{ruledtabular}\label{table1}
\begin{tabular}{ccccccddd}
&$a_s$ & $A_{10}$ & $A_1$ & $A_2$ & $\gamma$ \\
\hline
 Cu(111)& 3.94 & -0.4371 & 0.1889 & 0.1590 & 2.9416\\
  Cu(100)& 3.415 & -0.4218 & 0.2241 & 0.1389 & 2.5390\\
\end{tabular}
\end{ruledtabular}
\end{table}

To ensure the continuity of the pseudo potential in Eq. (\ref{Vion})  at the boundaries, we multiply it by a damping function of the form $[1 + \tanh(z)]/2$. The time-dependent interaction $H_I(t)$ in Eq. (\ref{tdse}) is described in the length gauge with the use of the dipole approximation
\begin{equation}\label{tdi}
H_I(t) = -z S(z) (F_{NIR}(t) + F_{IR}),
\end{equation}
where the function $S(z)$ accounts for screening of the electric field inside the Cu metal surface (see, e.g., Refs.~\cite{Kazansky2009,Aguirre2016}) and takes the form 
\begin{equation}\label{screen}
S(z) =  0.5\{1+\tanh[6z+3\xi)/\xi]\}.
\end{equation}
Here $\xi$ is the screening length. In Eq. (\ref{tdi}), $F_{NIR}(t)$ and $F_{IR}(t)$ describe, respectively, the NIR multi-cycle chirped pulses and IR single-cycle pulse. The chirped pulse has the form 
\begin{equation}\label{F1}
F_{NIR}(t) = E_{NIR}f(t)\cos(\omega_{NIR} t + \phi(t)).
\end{equation}
The function $f(t)$ is the pulse envelope and is chosen to be of a cosine square form, $\omega_{NIR}$ is the central angular frequency, and $E_{NIR}$ the strength of the laser pulse. The function $\phi(t)$ in Eq. (\ref{tdi}), which is a time-dependent carrier-envelope phase (CEP), determines the form of the chirped pulse. The latter has a hyperbolic form similar to the one used in Ref. \cite{Carrera2007}
\begin{equation}\label{chirp}
 \phi(t) = -\beta \tanh (\frac{t-t_0}{\tau}),
\end{equation}
and the corresponding instantaneous frequency is obtained according to
\begin{equation}\label{wt}
 \omega(t) =  \omega_{NIR} - \frac{d\phi(t)}{dt}=  \omega_{NIR} -\frac{\beta}{\tau} [\cosh^2(\frac{t-t_0}{\tau})]^{-1},
\end{equation}
where the chirp parameters $\beta$ (given in rad) and $\tau$ can be adjusted to control, respectively, the frequency sweeping range and the steepness of the chirped pulse centred at $t_0$. In Eqs. (\ref{chirp}) and (\ref{wt}) the optical phase $\phi(t)$ becomes zero in the case of $\beta=$ 0, and thus we have $\omega(t)=\omega_{NIR}$.

{The chirped pulse we use, although, can be considered idealist, it is motivated by the fast progress of the use of femtosecond-laser frequency combs \cite{Chang2022,Cundiff2003}, which consist of a regular comb of sharp lines. This class of laser sources has the advantage of providing precise control of spectroscopy of atoms, molecules and solid-state systems. We therefore believe that the recent achievements in optical frequency comb metrology \cite{Udem2002}, should help producing such chirped pulses in the near future.}

The IR single-cycle pulse has the form as in our previous work \label{Malek2021}
\begin{equation}\label{F2}
F_{IR}(t) = E_{IR}\mathrm{e}^{-t^{2}/(2\sigma_{IR}^{2})}15.53\frac{t}{\tau_{IR}},
\end{equation}
where $\sigma_{IR}=\tau_{IR}/4\sqrt{2\ln(2)}$ is the width of the Gaussian function in Eq. (\ref{F2}). Here, $\tau_{IR}=2\pi/\omega_{IR}$, and $\tau_{NIR}=2\pi/\omega_{NIR}$ are the total duration of the IR and NIR pulses. The parameters $\omega_{IR}$ ($\omega_{NIR}$) and $E_{IR}$ ($E_{NIR}$) are respectively, the central frequency and the strength of the IR pulse (NIR pulse). The amplitude $E_{i}$ $(i=NIR,IR)$ is related to the peak intensity via the relation $I_i=E_i^{2}$.

We calculate the HHG spectrum $H(\omega)$ by carrying out the Fourier transform of the expectation value of the dipole acceleration along the $z$-axis
\begin{equation}\label{H}
H(\omega) = |D_z(\omega)|^2,
\end{equation} 
where $D_z(\omega)$ is defined by
\begin{equation}\label{Dzw}
D_z(\omega) = \frac{1}{\sqrt{2\pi}} \int_{t_i}^{t_f} <D_z(t)> \mathrm{e}^{-i\omega t} dt,
\end{equation} 
and the time-dependent expectation value of the dipole acceleration $<D_z(t)>$ is written as
\begin{equation}\label{Dzt}
D_z(t) = \langle \psi(t)| \frac{\partial V(z)}{\partial z} + (F_{NIR}(t) + F_{IR}) |\psi(t)\rangle.
\end{equation}  

In Eq. (\ref{Dzw}) $t_i$ and $t_f$ define, respectively, the time at which the pulse is switched on and off. Note that the dipole acceleration $<D_z(t)>$ in Eq. (\ref{Dzt}) is convoluted by a window function of a Gaussian form $\exp{[-(t-t_0)^2/(2\sigma^2)]}$ centred at $t_0$ and having the width of $\sigma$=5.77/$\omega_{0}$. {The Gaussian filter has been widely used in time-frequency analysis to eliminate noises resulted from the use of the Fourier transform algorithm. Furthermore, it allows in general a faster decaying of a desired function at the boundaries, which in our case is the dipole accelerator, thus allowing to observe HHG plateaus with higher visibility.}

In our numerical simulations the initial states are obtained by diagonalizing the matrix representation of the Hamiltonian in Eq. (1) in the absence of the laser pulse, and which is constructed using of a sinus-DVR basis~\cite{Lill1982}. The time evolution of the electronic wave function $\psi(t)$, which satisfies the TDSE [cf. Eq. (\ref{tdse})], is solved numerically using a split-operator method combined with a fast Fourier transform (FFT) algorithm. The calculations are carried out in a grid of size $L_z$ = 8190 with the spacing grid $dz$ = 0.25 a.u., i.e. $n_z$=32768 grid points. The time step used in the simulation is $\delta t$=0.02 a.u.. The convergence is checked by performing additional calculations with twice the size of the box and a smaller time step. An absorbing boundary is employed to avoid artificial reflections, but without perturbing the inner part of the wave function. The boundary is chosen to span 10\% of the grid size in the $z$-direction.

\section{RESULTS AND DISCUSSION}\label{results}

\begin{figure}[h!]
\centering
\includegraphics[width=7.cm,height=4.5cm]{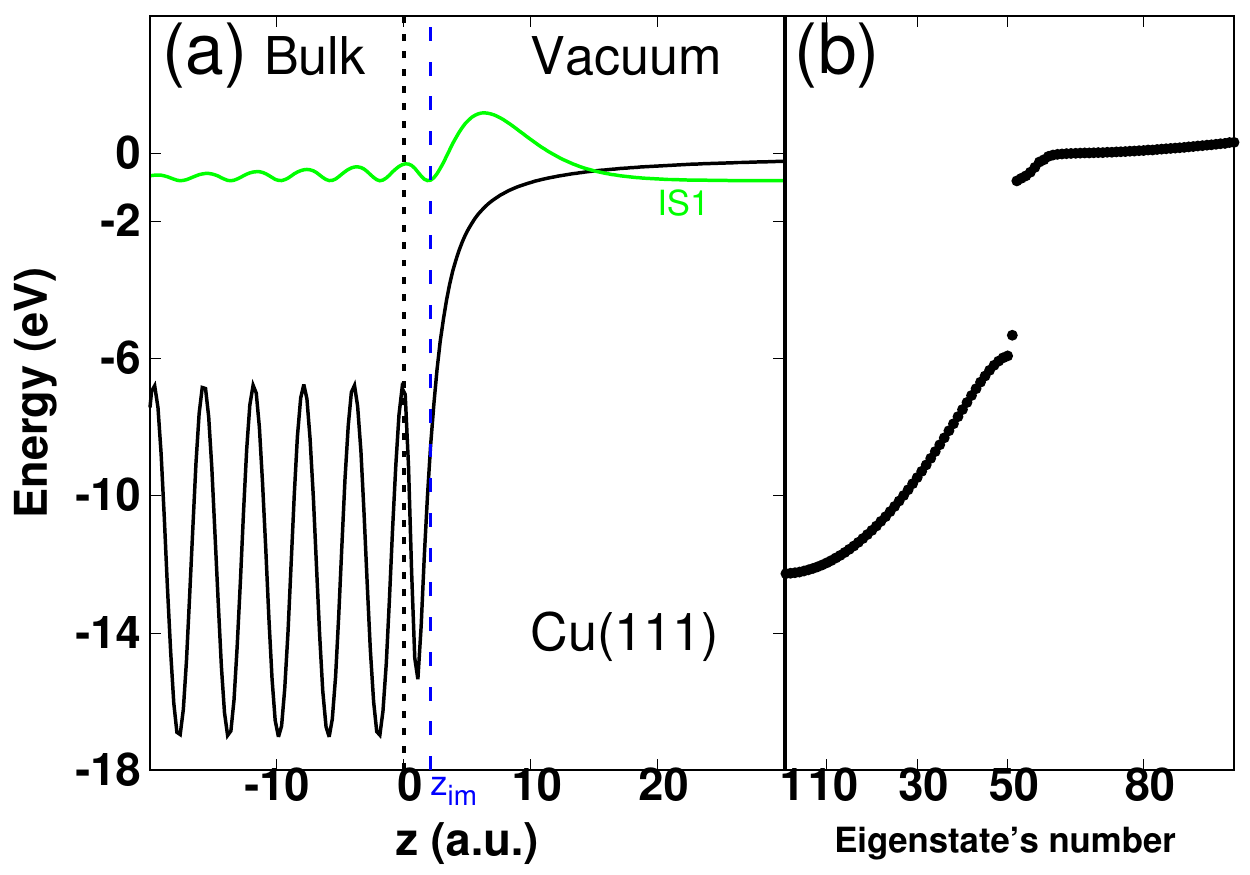}
\includegraphics[width=7.cm,height=4.5cm]{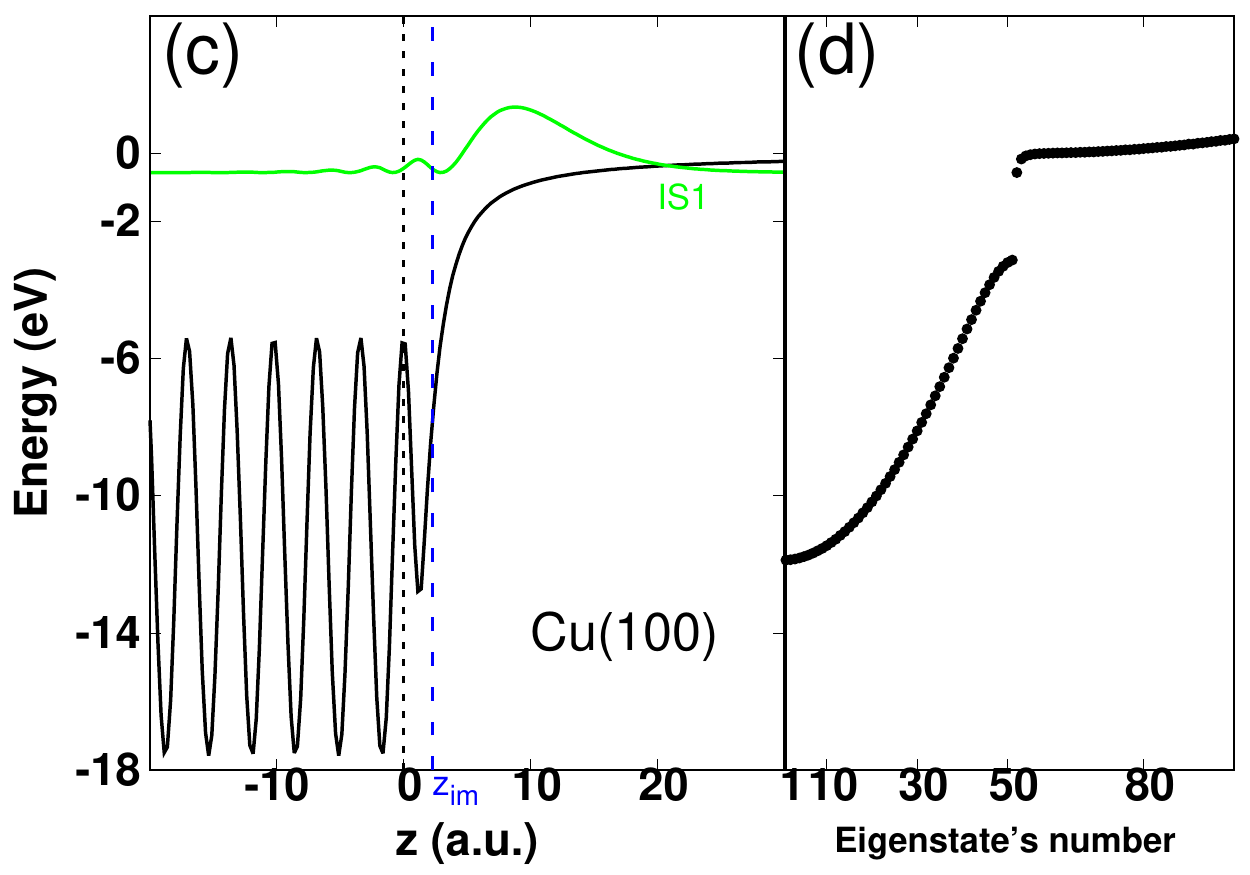}
\caption{\label{fig1} (Color online). Effective one-dimensional potential of Cu(111) (upper panel) and Cu(100) (lower panel) and the electron densities of the image-potential state having the energies -0.81 eV and -0.57 eV for Cu(111) and Cu(100), respectively. The corresponding electron densities are plotted (green (light gray) curve) with a shift from the origin to match the position of the state in the gap. (b) and (d) Eigenenergies at $k_{\parallel}=0$ of the binding electronic states of the model potential in (a) and (c). Here $k_{\parallel}$ is the crystal momentum of the electron. The pseudo potential is plotted in the limited range [-20:30] (a.u.) along the $z$-axis, and the derivative of the potential is continuous (see text).}
\end{figure}

\subsection{Model potential of Cu(111) and Cu(100)}

We consider the metal surfaces Cu(111) and Cu(100) to be initially prepared in an image-potential state. The latter arises from the interaction of an electron in front of the surface with its positive image charge in the bulk~\cite{Echenique1978,Klamroth2001}. The image states IS1 have the energies -0.81 and -0.57 eV, respectively, for Cu(111) and Cu(100), in accordance with the well-known formula of a hydrogen-like Rydberg series~\cite{Echenique1989} $-\frac{1}{16} \frac{1}{2(n_{im} + a)^2}$, where $n_{im}$ is the image-state index and $a$ is a quantum defect parameter of a surface of interest. This parameter is set to be 0.02 and 0.24 for Cu(111) and Cu(100), respectively~\cite{Chulkov1999,Klamroth2001}. The energies are located above the energy Fermi level -4.62 eV and close to the top of the  band gap as depicted in Figs. \ref{fig1}(b) and (d), and the corresponding electron densities are shown in Figs. \ref{fig1}(a) and (c). These electron densities exhibit a localized character of the states and are mainly confined in the metal-vacuum interface as their wavefunctions vanish towards the bulk as well as towards the vacuum~\cite{So2015}. A comparison between Cu(111) and Cu(100) shows that the maximum of the electron density of Cu(100) is shifted towards the vacuum by 2.5 a.u. compared to that of Cu(111). Consequently, its binding energy of the image-potential state is lowered.

\begin{figure}[h!]
\centering
\includegraphics[width=7cm,height=5.cm]{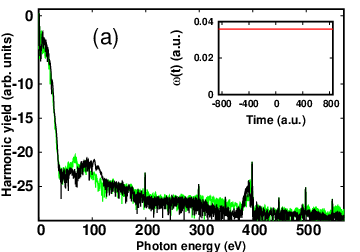}
\includegraphics[width=7cm,height=5.cm]{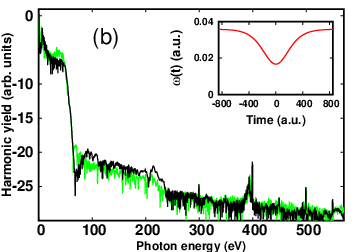}
\includegraphics[width=7cm,height=5.cm]{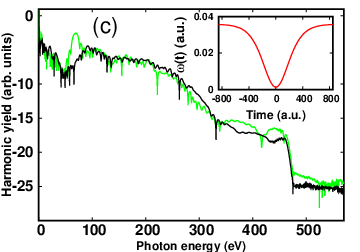}
\caption{\label{fig2} (Color online). HHG spectra induced by a chirped laser pulse alone calculated from the image-potential state in Cu(111) (green (light gray) curve) and Cu(100) (black curve) at different chirp parameter values: (a) $\beta=$ 0; (b) $\beta=$5; and (c) $\beta=$9 rad. Insets show the instantaneous frequency. The spectra are plotted in logarithmic scale. The parameters of the chirped pulse are: $\lambda_{NIR}=$ 1.27 $\mu m$, $T_c=$ 10 cycles, $\tau=$ 300 a.u. and $I_{NIR}=$ 2$\times$10$^{13}$ W/cm$^2$.}
\end{figure}

\subsection{HHG from image states in a single-color scheme}

\subsubsection{{Effect of the chirped pulse on HHG}}

We first consider a scenario involving single-color scheme in which a chirped laser pulse is linearly polarized along the [111] and [100] directions of the metal surface Cu. The pulse is characterized by a chirp parameter $\beta$ and has 1.27 $\mu m$ central wavelength, 43 fs pulse duration and 2$\times$10$^{13}$ W/cm$^2$ as the maximum of the peak intensity. 

Under the use of laser pulses, most of the damages are caused by plasma effects and ablation. Based on the model proposed in \cite{Gamaly2002}, a value of 0.51 J/cm$^2$ was set as the estimated damage threshold for copper. This value was found to be close to the experimental value 0.5-0.6 J/cm$^2$ (780 nm, pulse duration of 150 fs) \cite{Momma1997}. In our work, the laser parameters we use correspond to a fluence of 0.32 J/cm$^2$ \cite{Aguirre2016}, which is too low to cause any damage of the material. We therefore expect that the laser parameters we consider in this work can barely destroy the sample.

We start our discussion by computing HHG spectra for $\beta=$0, 5 and 9 rad for both surface orientations [111] (green curve) and [100] (black curve) as shown in Fig. \ref{fig2}. The results show a strong sensitivity of HHG spectra to the change of the chirp parameter as well as to the orientations . In particular, it is seen that the cutoff harmonic extends from $E_c=$25 eV for $\beta=$0 (free-chirp case) to $E_c=$450 eV for $\beta=$9 rad. An extension by almost a factor of 18 is observed. This significant extension is linked to the high-kinetic energy that the free electrons acquired from the field~\cite{Agueny2021}. In a classical picture, this energy is referred to as the ponderomotive energy of free electrons in an oscillating field (i.e. $U_p=I/4\omega^2$) and is inversely proportional to the instantaneous frequency characterized by the chirp parameter $\beta$. The instantaneous frequency is shown in Fig. \ref{fig2} as inset. Here one can see that increasing the parameter $\beta$ results in decreasing of the frequency $\omega(t)$, and hence a vast kinetic energy $U_p$ is transferred to free electrons, which results in an extension of the cutoff energy. A similar picture is known in HHG from atomic gases (e.g.~\cite{Carrera2007}). These results on the other hand, indicate that shaping the laser pulses by changing the chirp parameter has the advantage of amplifying the acquired energy from the laser fields. 

\begin{figure*}[ht]
\centering
\includegraphics[width=7cm,height=5.cm]{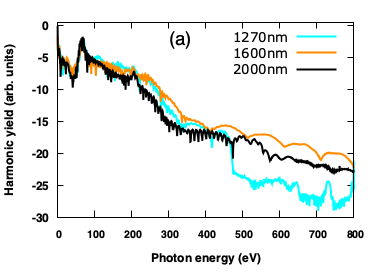}
\includegraphics[width=7cm,height=5.cm]{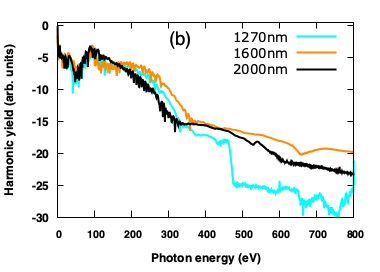}
\includegraphics[width=7cm,height=5.cm]{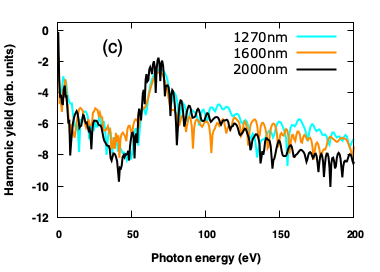}
\includegraphics[width=7cm,height=5.cm]{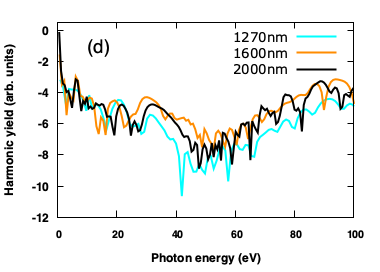}
\caption{\label{fig21} {(Color online). HHG spectra induced by a chirped laser pulse alone calculated from the image-potential state in Cu(111) (a) and Cu(100) (b) at different wavelengths: $\lambda_{NIR}=$ 1.27 $\mu m$ (cyan (light gray) curve), $\lambda_{NIR}=$ 1.6 $\mu m$ (orange curve), and $\lambda_{NIR}=$ 2.0 $\mu m$ (black curve). A zoom of the spectra in (a) and (b) is shown in (c) and (d), respectively. The spectra are plotted in logarithmic scale. The parameters of the chirped pulse are: $\beta=$9 rad, $T_c=$ 10 cycles, $\tau=$ 300 a.u. and $I_{NIR}=$ 2$\times$10$^{13}$ W/cm$^2$.}}
\end{figure*}

A comparison of HHG spectra between the surface orientations [111] and [100] shows an enhancement of the harmonic yield in the case of Cu(111) by almost one order of magnitude in the photon energy range 50-100 eV. This energy range corresponds to the XUV photon energy window, and that is an interesting finding for generating XUV IAPs with higher flux intensities in an experiment. A similar enhancement has been seen in HHG from the surface state in Cu(111) and was explained in terms of quantum-path interference. The emergence of these quantum effects in our calculations from an image-potential state and their absence in the Cu(100) case reveal the importance of trapped electrons in metal surfaces for generating and characterizing IAPs in the XUV and soft x-ray regions. The observed enhancement is not limited to the wavelength 1.27 $\mu m$, but it extends to laser wavelengths in the spectral range 1-2 $\mu m$. In this range, the above discussion remains valid. This is shown in Figs. \ref{fig21}(a) and \ref{fig21}(b), in which we present HHG spectra for 1.6 and 2.0 $\mu m$ in both orientations Cu(111) and Cu(100). 

A closer inspection of the spectra at the low photon energy (cf. Figs. \ref{fig21}(c) and \ref{fig21}(d)) reveals some similarities in the behavior of HHG for the wavelengths considered here. This behavior is found to be sensitive to the surface orientation of the metal. The investigation of its origin is however beyond the scope of the present work, which calls for additional works to shed new light on this observed behavior. On the other hand, one can see that beyond the photon energy 400 eV the spectra exhibit a fast oscillatory structure. In order to rule out their origin to be caused by numerical noise, we have performed calculations with a larger spatial grid of $L_z=$16380 a.u. ($n_z=$65536 points) and a smaller time step (i.e $\Delta=$0.02 a.u.), the results are found to be indistinguishable as shown in Supplemental Material~\cite{SM}, thus ensuring their convergence. Furthermore, we have observed that beyond the laser wavelength 2 $\mu m$, the HHG efficiency decreases similarly to the case of gases (see Supplemental Material~\cite{SM}).

\subsubsection{{Effect of the peak intensity on HHG}}

\begin{figure}[h!]
\centering
\includegraphics[width=7cm,height=5.cm]{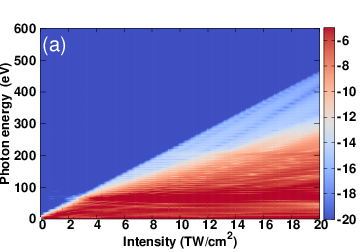}
\includegraphics[width=7cm,height=5.cm]{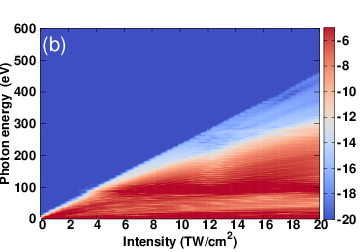}
\caption{\label{fig222} {(Color online). Two-dimensional HHG spectra induced by a chirped laser pulse at different peak intensities in the range [0.5-20] TW/cm$^2$. The spectra are calculated from the image-potential state in Cu(111) (a) and Cu(100) (b). The spectra are plotted in logarithmic scale and are given in arb. units. The parameters of the chirped pulse are: $\beta=$9 rad, $\lambda_{NIR}=$ 1.27 $\mu m$, $T_c=$ 10 cycles, $\tau=$ 300 a.u..}}
\end{figure}

Additional calculations illustrating the effect of varying the peak intensity of the chirped pulse on the HHG spectra are displayed in Fig. \ref{fig222}. We use the same parameters as in Fig. \ref{fig2}(c). The results are shown for both surface orientations [111] and [100] respectively, in Figs. \ref{fig222}(a), \ref{fig222}(b). As expected, the photon energy cutoff increases linearly as a function of the peak intensity and follows the approximative formula
\begin{equation}\label{Emax}
E_{max}(\beta) = I_p + 3.17Up(\beta),
\end{equation} 
where $I_p$ is the ionization potential. The formula (\ref{Emax}) deviates from the well known one due to its dependence on the chirp parameter $\beta$ via the instantaneous frequency $\omega(t)$. It was found however that this formula reproduces very well the energy cutoff by including some adjustments to take into account the instantaneous variation of the kinetic energy $U_p$ \cite{Agueny2021}.

In Fig. \ref{fig222}, it is seen that the linear behavior of the energy cutoff is found to be sensitive to the surface orientations Cu(111) and Cu(100). This is expected as the initial state exhibits some differences in both the bulk and vacuum regions as described above [cf. Figs. \ref{fig1}(a) and \ref{fig1}(c)], which in turn modify the HHG spectra. We stress here that the observed linear scaling has been already observed in the chirp-free case \cite{Aguirre2016} and found to be caused by the electron traveling through energy dispersion bands \cite{Vampa2015c}.

\subsubsection{{Effect of the pulse duration on HHG}}

\begin{figure}[h!]
\centering
\includegraphics[width=7cm,height=5.cm]{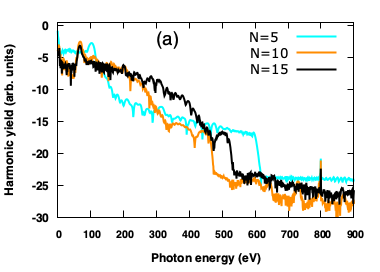}
\includegraphics[width=7cm,height=5.cm]{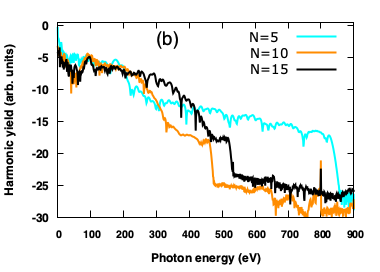}
\caption{\label{fig22} (Color online). HHG spectra induced by a chirped laser pulse alone calculated from the image-potential state in Cu(111) (a) and Cu(100) (b) at different number of cycles: $N=5$ cycles (cyan (light gray) curve), $N=10$ cycles (orange curve), and $N=15$ cycles (black curve). The spectra are plotted in logarithmic scale. The parameters of the chirped pulse are: $\beta=$9 rad, $\lambda_{NIR}=$ 1.27 $\mu m$, $T_c=$ 10 cycles, $\tau=$ 300 a.u. and $I_{NIR}=$ 2$\times$10$^{13}$ W/cm$^2$.}
\end{figure}

In order to investigate how the pulse duration modifies the HHG spectra, we perform calculations at two different durations of the chirped pulse for both Cu(111) [cf. Fig\ref{fig22}(a)] and Cu(100) [cf. Fig\ref{fig22}(b)]: 5-cycles (cyan (light gray) curve) and 15-cycles (black curve). For reference, the spectra obtained with 10-cycle pulses are also shown in the same figure with orange color. These pulses correspond to a duration of 21.18, 42.36 and 63.54 fs, respectively for 5, 10 and 15-cycle pulses. The results are shown for the laser wavelength 1.27 $\mu m$ and at the peak intensity of 2 TW/cm$^2$. The chirp parameter is fixed at $\beta=9$. It is seen that changing the pulse duration modifies dramatically the harmonic yield as well as the photon energy cutoff. All the spectra exhibit an interesting signature that reflects the dynamics behind the change of the pulse duration. It is found that the shorter pulse with 5-cycle induces a substantial extension of the photon energy cutoff, in particular for the surface orientation [100], in which higher photon energies up to 850 eV are generated. A similar behavior has been observed for the Cu(111)-NaCl system using a chirp-free pulse \cite{Aguirre2016}. On the other hand, an enhancement of the harmonic yield by two order of magnitude is observed at the photon energy 100 eV for the geometry [111]. Whereas, the use of a longer pulse with 15-cycles is not optimal for producing interesting results when compared with the spectra for 5 and 10-cycle pulses.

\begin{figure*}[ht]
\centering
\includegraphics[width=7cm,height=6.cm]{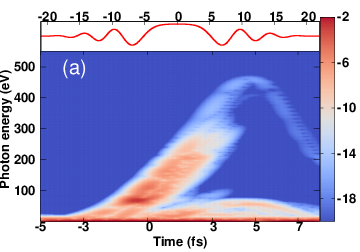}
\includegraphics[width=7cm,height=6.cm]{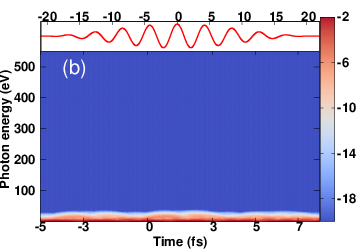}
\includegraphics[width=7cm,height=6.cm]{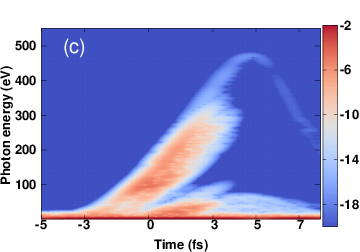}
\includegraphics[width=7cm,height=6.cm]{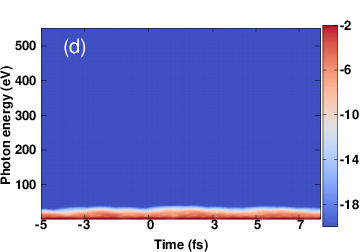}
\includegraphics[width=7cm,height=6.cm]{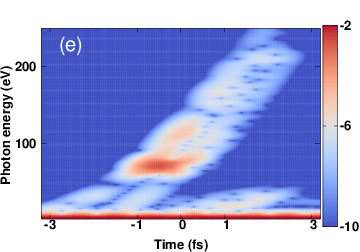}
\includegraphics[width=7cm,height=6.cm]{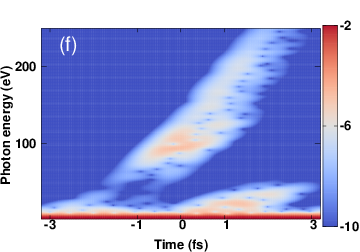}
\caption{\label{fig3} (Color online). Time-frequency analysis of the harmonic spectra for Cu(111) (upper panel) and Cu(100) (middle panel). A zoom of the spectra in (a) and (b) in the photon energy [1-250] eV and time interval [-3:3] fs is shown in the lower panel for Cu(111) (e) and Cu(100) (f). For reference, the spectra for $\beta=$0 for Cu(111) (b) and Cu(100) (d) are also shown. Insets: the form of the chirped pulse (a) and free-chirped pulse (b). The spectra are plotted in logarithmic scale and are given in arb. units. The parameters of the chirped pulse are: $\lambda_{NIR}=$ 1.27 $\mu m$, $T_c=$ 10 cycles, $\tau=$ 300 a.u. and $I_{NIR}=$ 2$\times$10$^{13}$ W/cm$^2$.}
\end{figure*}

\subsubsection{{Time-frequency analysis of HHG}}

In the following our results will be presented for the laser wavelength 1.27 $\mu m$ and at the peak intensity of 2 TW/cm$^2$. Here, we adopt a Gabor time-frequency profile of HHG and extract amplitudes of harmonics in the XUV and soft x-ray regions. This Gabor-based analysis is widely performed in connection with the HHG process in atomic and molecular gases as well as solids. It has the advantage of providing detailed information about the time emission of the bursts and thus allowing to access the width of the generated attosecond pulses. We thus perform this analysis for both orientations Cu(111) and Cu(100) in the case of a chirped [cf. Fig. \ref{fig3}(a) and (c)] and unchirped pulses [cf. Fig. \ref{fig3}(b) and (d)]. The obtained results are displayed for a peak intensity of 2$\times$10$^{13}$ W/cm$^2$ and for a fixed chirp parameter of $\beta=$9 rad. The figure provides a comparative overview of the role of chirped pulses as well as the surface orientation for spectral characterizations of emission bursts.

In particular, one can see the emission of a pronounced burst in both orientations of Cu, while it is absent in the free-chirped case. The emission bursts extend to higher energies and up to 450 eV when shaping the laser pulses and are mostly generated in the time interval [-3:3] fs. The emitted bursts look similar for both surface orientations. A closer inspection of the spectra however reveals some distinct features that emerge with higher visibility in the time interval [-1.5:0.5] fs as shown in Figs. \ref{fig3}(e) and (f) for Cu(111) and Cu(100), respectively. These features encode information about the electronic band structure of the metal surface having a specific orientation, which in turn can be mapped out in the generated IAPs.

\subsection{Isolated attosecond pulses in a two-color scheme}

\begin{figure*}[ht]
\centering
\includegraphics[width=7cm,height=5.cm]{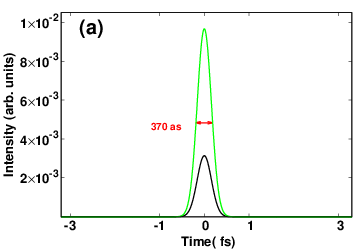}
\includegraphics[width=7cm,height=5.cm]{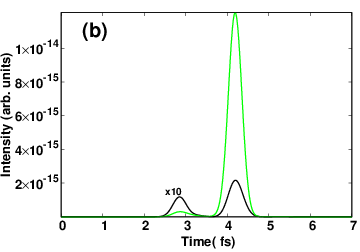}
\includegraphics[width=7cm,height=5.cm]{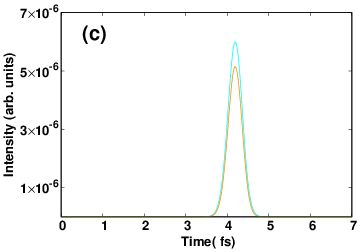}
\includegraphics[width=7cm,height=5.cm]{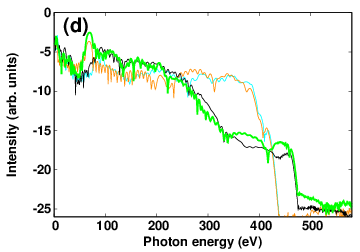}
\caption{\label{fig4} (Color online). (upper panel) Intensity of the extracted IAPs with the chirped pulse alone for Cu(111) (green (light gray) curve) and Cu(100) (black curve) at two different window energies: (a) 50-250 eV and (b) 350-450 eV. The amplitude of IAP for Cu(100) in (b) is multiplied by 10 to make it visible. The width of the IAPs is indicated in the figure (a). (c) Intensity of IAPs in the presence of a single-cycle for Cu(111) (orange curve) and Cu(100) (cyan (light gray) curve). (d) The corresponding HHG spectra displayed with the same colors as in (a) and (c). The green curve in (d) is displayed with a thick line. The parameters of the chirped pulse are: $\lambda_{NIR}=$ 1.27 $\mu m$, $T_c=$ 10 cycles, $\beta=$ 9rad and $I_{NIR}=$ 2$\times$10$^{13}$ W/cm$^2$. The parameters of the IR single-cycle pulse are: $\lambda_{IR}=$ 12.7 $\mu m$ and $I_{IR}=$ 1$\times$10$^{12}$ W/cm$^2$.}
\end{figure*}

Based on the Gabor analysis displayed in Fig. \ref{fig3}, we extract the intensities of the harmonics covering a broad window energy as defined below. Specifically, we compute a coherent sum of the consecutive harmonics in a well-defined energy region according to~\cite{Mairesse2003,Lan2006}      
\begin{equation}\label{IAP}
I_z(\tau) = | \frac{1}{\sqrt{2\pi}}  \sum_{\omega_i}^{\omega_f}  \int_{t_i}^{t_f} \exp{[-\frac{(t-\tau)^2}{2\sigma^2}]}<D_z(t)> \mathrm{e}^{-i\omega t} dt|^2,
\end{equation} 
where $\omega_i$ and $\omega_f$ represent the energy window used for extracting the intensity of IAPs. The results are presented in Fig. \ref{fig4} and display the extracted IAPs having a broad bandwidth covering the energy window 50-250 eV [cf. Fig. \ref{fig4}(a)] and 350-450 eV [cf. Fig. \ref{fig4}(b)]. These regions are chosen to cover regular high-energy plateaus emerging in HHG spectra as shown in Fig. \ref{fig2}(c). Here, the duration of the extracted IAPs is found to be about 370 as for both surface orientations [111] and [100]. However, the intensity of IAP is one order of magnitude higher in the [111] orientation case. An extension of the extraction procedure to a broader energy window (i.e. 350-450 eV) shows the appearance of a burst with a weak intensity [cf. \ref{fig4}(b)]. 

In order to enhance its intensity we introduce an additional IR single-cycle pulse with a peak intensity of $1\times$10$^{12}$ W/cm$^2$ and central wavelength of $\lambda_{IR}=10\lambda_{NIR}$. {The use of such pulses is motivated by the recent work \cite{Nie2018} in which the authors demonstrate the possibility of producing IR single-cycle pulses covering the spectral range 5-14 $\mu m$ and with the intensity of the order of terawatt per cm$^2$.}

The basic physics behind these single-cycle pulses is related to a high-momentum transfer to electrons~\cite{Sha2014,Yang2014,Chovancova2017}, which results in a high-energy recollision, and hence extension of the harmonic cutoff and enhancement of the harmonic yields as it has been discussed recently in the context of an atomic system~\cite{Malek2021}. The effect of introducing an IR single-cycle pulse on HHG can be seen in Fig. \ref{fig4}(d). Interesting enough, an enhancement of the harmonic yield by almost seven order of magnitude is seen in both surface orientations Cu(111) (orange curve) and Cu(100) (cyan (light gray) curve). For reference, the HHG spectra obtained with the chirped pulse alone are also shown in the same figure. This enhancement is reflected in the extracted IAPs as one can see in Fig \ref{fig4}(c). These IAPs are generated with roughly the same intensity, unlike in the case with only the chirped pulses. This result elucidates the effect of introducing an IR single-cycle pulse to enhance the intensity of the IAPs in the soft x-ray photon energy region. Generating IAPs in this region is highly desirable for time-resolved measurements with attosecond precision.

At this point we conclude that the localized aspect of image-potential states combined with tuning the optical properties of chirped pulses and the use of a single-cycle pulse are found to be relevant for generating ultrabroad IAPs with larger bandwidths. Although the role of localized states has been addressed previously, it has been done by creating a vacancy in a periodic system that is initially prepared in bulk states (delocalized states) \cite{Mrudul2020}. Our work, however addresses this role from a different perspective: the dynamics is initiated from a localized state (image-potential state) in a scenario involving two different geometries (Cu(111) and Cu(100)), which leave their imprint in the HHG spectra. We note that such geometrical dependence in HHG has been reported previously in ferromagnetic monolayers \cite{George2018}. Our work thus complements the existing studies and adds new insights into the role of localized states in solid-state HHG.

With the state-of-the-art laser technology, it should be possible to generate IR single-cycle pulses with petawatt peak intensities and in the spectral range considered in our present work. For instance, a new scheme capable of producing relativistic single-cycle pulses of wavelength-tunable (5‚Äì14 $\mu m$) has been proposed ~\cite{Nie2018,Zhu2020,Ma2012}. On the other hand, producing chirped femtosecond pulses should be possible using the current state-of-the-art techniques, which benefit from the nowadays ultrafast laser technologies (e.g.~\cite{Gauthier2016}). A widely used technique that is considered as a standard laser configuration is the chirped pulse amplification~\cite{Strickland1985}, which is based on the stretching and compression of the laser pulse. In this technique, the control of the chirped pulse can be achieved using for instance active devices such as deformable mirrors or liquid crystal arrays~\cite{Kartner2001,Hong2002}. On the other hand, generating positive chirp, as in our work, can be done in an experiment using an imperfect dispersion compensation~\cite{Hong2002}, which can be achieved by tuning either a pulse stretcher or a pulse compressor. Furthermore, the characterization of the chirped pulse can be realised using frequency-resolved optical gating technique (FROG)~\cite{Clement1995}. We thus believe that the two-color scheme proposed in our work to generate XUV and soft x-ray IAPs should be experimentally feasible in the near future. {On the other hand, despite the difficulties in realizing an experiment of the HHG process from metal surfaces \cite{Ghimire2014,Korobenko2021}, our work provides new insights into their role for coherent control of the process. It thus functions as a benchmark for future studies of solid-state HHG and in general of strong-field light-matter interaction.} 

\section{CONCLUSIONS}\label{conclusions}

In conclusion, we have demonstrated the relevance of trapped electrons in metal surfaces for generating ultrabroad IAPs when they are excited by means of optimized chirped pulses combined with an IR single-cycle pulse. In particular, we have predicted the generation of IAPs with a duration of about 370 as and with a bandwidth covering the photon energy of 50-250 eV and 350-450 eV. The generated IAPs were found to be sensitive to the surface orientations Cu(111) and Cu(100), which suggest their use for characterizing the IAPs. Moreover, we have shown the effect of introducing an IR single-cycle pulse for enhancing the intensity of the extracted soft x-ray IAPs. We have found an enhancement by a factor of seven order of magnitude compared to the case with the chirped pulse alone. Our findings thus indicate the relevance of trapped electron excited by a two-color scheme for extending attosecond metrology to the XUV and soft X-ray regions. This extension is critical towards harnessing attosecond quantum technologies from solid-state systems.  


\section*{ACKNOWLEDGMENTS}
The author AY acknowledges support from SPIRA during his stay at the University of Bergen. The author A.T acknowledges support from the bilateral relationships between Morocco and Hungary in Science and Technology (S \& T) under the project number 2018-2.1.10-TAâT-MC-2018-00008.

\providecommand{\noopsort}[1]{}\providecommand{\singleletter}[1]{#1}%

\end{document}